\documentclass[12pt]{article}

\usepackage{natbib}

\usepackage[colorinlistoftodos]{todonotes}
\usepackage{arxiv}
 
\usepackage{setspace}

\usepackage[utf8]{inputenc} 
\usepackage[T1]{fontenc}    
\usepackage{hyperref}       
\usepackage{url}            
\usepackage{booktabs}       
\usepackage{amsfonts}       
\usepackage{nicefrac}       
\usepackage{microtype}      
\usepackage{lipsum}

\usepackage{caption}
\usepackage{graphicx}
\usepackage{lscape}
\usepackage{hyperref}
\usepackage{amsmath}
\usepackage{longtable}%
\usepackage{colortbl}%

\title{Making Markets for Information Security:\\ The Role of Online Platforms in Bug Bounty Programs}

\author{
  Johannes Wachs\\
  Vienna University of Economics and Business\\
  Complexity Science Hub Vienna\\
  johannes.wachs@wu.ac.at
}

\begin{document}
\maketitle

\begin{abstract}

Security is an essential cornerstone of functioning digital marketplaces and communities. If users doubt that data shared online will remain secure, they will withdraw from platforms. Even when firms take these risks seriously, security expertise is expensive and vulnerabilities are diverse in nature. Increasingly, firms and governments are turning to bug bounty programs (BBPs) to crowdsource their cybersecurity, in which they pay individuals for reporting vulnerabilities in their systems. And while the use of BBPs has grown significantly in recent years, research on the actors in this market and their incentives remains limited. Using the lens of transaction cost economics, this paper examines the incentives of firms and researchers (sometimes called hackers) participating in BBPs. We study the crucial role that centralized platforms that organize BBPs play in this emerging market. We carry out an analysis of the HackerOne BBP platform, using a novel dataset on over 14,000 researchers reporting over 125,000 public vulnerabilities to over 500 firms from 2014 to the end of 2021. We outline how platforms like HackerOne make a market for information security vulnerabilities by reducing information asymmetries and their associated transaction costs.\\

\textit{Keywords:} Online Markets, Information Security, Crowdsourcing, Trust, Hacking
\end{abstract}

\onehalfspacing
\section*{Introduction}
Information security plays a key role in the digital economy, especially on sites and platforms where users provide private information or exchange money for goods and services. These online platforms seem to work best when they establish trust with their users, meaning that users feel secure because of features of the system the platform operates, rather than the behavior of other users \citep{pavlou2004building}. As more users participate in these online platforms and marketplaces, the security stakes rise and systems are coming under increasing pressure from malicious actors that seek to steal information or defraud users, leading to a variety of harmful outcomes \citep{agrafiotis2018taxonomy}. And online platforms are not the only domain in which information security has become a major issue. Indeed, as digitization extends to a wider variety of consumer goods and services from automobiles \citep{svangren2017connected} to e-government \citep{zhao2010opportunities}, the scope of potential vulnerabilities has only increased. Vulnerabilities often have externalities, affecting other firms and individuals beyond the confines of a single firm \citep{anderson2001information,gordon2014externalities,ahmed2021modelling}, suggesting that investment in information security will be suboptimal from a social welfare perspective \citep{gordon2015impact}. Digital insecurity is known to have disparate impact across socio-economic groups, compounding disadvantages for certain groups \citep{hargittai2021handbook}. Hacking has even become a geopolitical issue, as state actors are both victims and perpetrators of attacks \citep{buchanan2020hacker}.

How do firms and institutions organize their defense? Traditional in-house information security functions are often supplemented by consultants who carry out security audits using techniques such as penetration testing. However, the growing complexity of software systems makes it difficult for firms to scope their security needs and for experts to anticipate the many avenues of potential attacks \citep{biro2017software,malladi2019bug}. Given large and growing demand for this kind of knowledge, it has become increasingly expensive to hire security experts \citep{libicki2014hackers}. 

One alternative attracting attention in recent years is the concept of a bug bounty program (BBP). Bug bounties are a way of crowdsourcing security in which a firm pledges to reward outside researchers (colloquially referred to as hackers) with financial bounties for reporting vulnerabilities they observe in the firm's systems. By stating what kind of vulnerability reports they are interested in and how they handle and potentially reward such reports, firms can signal to the community of security researchers that they are open to such exchanges. The core idea is that BBPs, when well implemented, offer a kind of security audit that harnesses the knowledge and know-how of a much larger and more diverse group of individuals than any in-house security team or consulting firm can offer \citep{maillart2017given}.

Historically, BBPs emerged in a decentralized manner, with firms and institutions offering bounties and setting up rules of engagement on their own \citep{goerzen2022wearing}. And while the decentralized market for information security vulnerabilities seemed to grow significantly in the late 90's into the 00's, several issues limited such exchanges between firms and researchers. For example, a firm may already be aware of a specific vulnerability when a researcher reports it, and then be unwilling to reward that researcher. A firm could also underpay a researcher relative to the value of the exposed vulnerability, especially if they think that this is a one-off scenario. Even if both parties have good intentions, the inherent complexity of vulnerability reports makes it difficult to coordinate their resolution and valuation. 

Researchers familiar with Transaction Cost Economics (TCE) \citep{williamson1979transaction} will recognize these as examples of situations of asymmetric information and opportunistic behavior in markets. Moreover, as vulnerabilities are highly specific, complex, and of uncertain frequency, the theory of TCE suggests that information security will likely be organized within firms, as part of a hierarchy rather than in a market. Add to these observations the fact that vulnerabilities, if publicized, can come with significant regulatory and reputational costs \citep{goel2009estimating}, it is no wonder that firms often keep security functions in-house or simply try to ignore the problem. Similarly, researchers also face several obstacles when participating in bounty programs. For example, researchers value the publication of their report after the vulnerability has been resolved as a token of prestige and a credible demonstration of their skills, which carries significant value on the labor market \citep{goerzen2022wearing,ellis2022bounty}. A priori it is unclear which firms will agree to this kind of transparency and under what conditions, and even if a firm will pay a reasonable reward for such a report.

Overall it seems that information asymmetries are major obstacles to a well-functioning and large scale market for vulnerabilities. However, the same forces of digitalization that have increased the importance of information security in general have also introduced new technologies and organizational forms for coordinating economic activity. The Electronic Markets Hypothesis (EMH) of Malone, Yates and Benjamin \citeyearpar{malone1987electronic} predicted that digitalization would decrease transaction costs in general and expand the scope of activities that take place in markets, often coordinated by ``market makers''. We argue that this move to the market in information security has taken place in the last decade, lead by online platforms \citep{alt2019electronic} (the market makers) such as HackerOne and BugCrowd. As intermediaries, platforms decrease transaction costs that would otherwise prevent markets from being viable \citep{alt2017electronic,nagle2020transaction}. Platforms managing BBPs facilitate this market by reducing information asymmetries using standardization and reputation systems.

In this article we analyze the market for bug bounties through the perspective of Transaction Cost Economics, arguing that, in line with the Electronic Markets Hypothesis, centralized online platforms play a key role in making such markets possible by reducing information asymmetries. We give an overview of bug bounty programs and the incentives in this market from both the firm and researcher perspectives. We then suggest how a centralized platform can reduce these costs to the extent needed to make a viable market. Using data from HackerOne, a market leading BBP platform, we present empirical evidence connecting TCE theory with firm and researcher behavior. We find that researchers and firms use platform functionalities to solve information problems (i.e. to signal) and to better capture value of their activity. We also observe rising specialization of researchers in specific firms over time, suggesting the emergence of embedded structures in the market. We conclude with a summary of our contribution, and suggest some open questions in this area.

\section*{Background and Theoretical Framing}

In this section we provide an overview of BBPs, including a brief history. We highlight information problems that firms and researchers face when participating in bug bounties. We then theorize how centralized platforms for managing BBPs help resolve these problems. We then develop a set of hypotheses about these theories, which we can be tested using empirical data we have collected from the HackerOne platform.

\subsection*{Background on Bug Bounty Programs}

The first BBPs were established, perhaps unsurprisingly, by firms and institutions working in the software industry \citep{goerzen2022wearing}. Besides the fact that software is the natural arena of digital security, the industry has a long tradition of decentralized, crowd-based collaboration \citep{latoza2015crowdsourcing}, for instance in open source software (OSS). Indeed, recent work has adapted Eric Raymond's maxim on the effectiveness of OSS development, that ``with enough eyeballs, all bugs are shallow'' \citep{raymond1999cathedral} to the context of information security vulnerabilities via BBPs \citep{maillart2017given}. In this framing, BBPs are an organizational framework to crowdsource an important problem, and crowdsourcing itself is an effective way to harness diverse human resources and collective intelligence \citep{agrawal2014some,riedl2021quantifying,wachs2021does}. 

The first such programs emerged out of a volatile period in the information security community in the 1990s \citep{goerzen2022wearing}. Researchers and hackers investigating widely used software for vulnerabilities became increasingly frustrated with how firms responded to their findings either by ignoring them or reporting them to the authorities. The Bugtraq mailing list, started in 1993, became a place to fully disclose and document such vulnerabilities. Goerzen and Coleman \citeyearpar{goerzen2022wearing} argue that researchers and hackers operating outside institutional environments began participating in this mailing list to establish legitimacy for themselves and their community, and to exchange ideas with insiders. Though highly controversial at first, collective efforts like Bugtraq demonstrated both the effectiveness of crowdsourced insights into security and that many ``hackers'' could be trustworthy contributors. 

In parallel, leading tech firms established the first BBPs \citep{ellis2022bounty}. The term bug bounty program was coined at Netscape in 1995, when the company launched such a program with an initial budget of \$50,000. However, one condition of Netscape's BBP was that rewarded vulnerabilities would not be publicly disclosed. Indeed, as Ellis and Stevens \citeyearpar{ellis2022bounty} document, Netscape had experienced bad publicity when several vulnerabilities in its browser were published to Bugtraq. In this way, Netscape was attempting to limit the uncertainty and potential costs of vulnerabilities: a bounty paid to a researcher was likely cheaper in the long-term than a disclosed major vulnerability on Bugtraq. This highlights a major tension in BBPs: researchers often want a report to be disclosed for a variety of reasons, while firms tend to want to keep them secret.

It took some time for BBPs to become more widely adopted. Google and Facebook launched their own programs only in 2010 and 2011, respectively \citep{ellis2022bounty}. And while by the 2020s some companies host their own BBPs, the idea became really widespread, and used by smaller firms, only after the establishment of centralized platforms managing BBPs like HackerOne and Bugcrowd. Among private or invitation-only BBPs, NDA agreements are common and disclosure is rare. Activity on these centralized platforms is signficant: HackerOne reported that firms had paid out over \$36 million in bounties in 2021 \citep{hackerone2021}, with the median ``critical'' bug netting a researcher around \$3,000.

\subsubsection*{Researcher Participation in BBPs}
Though a full investigation of why individuals participate in BBPs as researchers is beyond the scope of this article, we briefly consider some motivations for participation here. Our emphasis is on how motivations come into conflict with the those of firms on the other side of a potential transaction or how lack of information may be a barrier to continued participation.

Around 40\% of researchers participating in a 2019 HackerOne survey claim they participate in BBPs in order to learn and build a career in information security \citep{hackerone2019}. Roughly the same share of respondents say they hack for fun (13\%) as for the money (14\%) - though this survey is not based on a representative sample. HackerOne materials emphasize that BBPs offer flexible work opportunities, where individuals can decide how much they work and with which companies they work. This and other aspects of hacking closely resemble the characterization of ``gig work'' or the ``gig economy'' \citep{graham2017digital,ellis2022bounty}. Many researchers do work only part time on BBPs \citep{hata2017understanding}, either supplementing their main income or as part of a transitory stage in their careers.

In practice researchers face several issues resulting from a lack of information. For example, a vulnerability that has already been reported but not yet fixed is rarely rewarded \citep{zhao2017devising}, creating a significant amount of wasted effort. While firms attempt to pre-commit to certain kinds of payments depending on the kind and severity of the vulnerability, many vulnerabilities are unique in some way and often quite complex. This makes it unclear how a vulnerability will be mapped to the menu of rewards, and can lead to a researcher being disappointed in their reward. Researchers can exchange information about how well firms treat them on message boards and community forums, but this introduces significant searching costs.

Finally, as discussed above, researchers are often interested in getting their vulnerability reports disclosed so that they can demonstrate their accomplishments. When building a career and resume in the information security sector, a credible record of having hacked well-known tech companies can go a long way \citep{ellis2022bounty}. These kinds of credentials can substitute for formal degrees, and may be even more valuable for individuals from developing countries. Firms can often simply refuse to disclose a report, and may be more likely to do so if the vulnerability is embarrassing or hints at other potential vulnerabilities. Disclosure is important from a more general perspective as well. Goerzen and Coleman highlight several reasons \citeyearpar{goerzen2022wearing}:

\begin{itemize}
    \item Disclosure opens knowledge about a vulnerability to a wide audience.
    \item Repeated disclosure creates a space for both sides to share information and arrive to common standards which can improve communication in subsequent interactions \citep{blind2016impact}.
    \item It resolve issues of credit and ``ownership'' of vulnerabilities.
    \item Competition among researchers creates a way to define reputation and to build CVs.
    \item Good examples demonstrate the social value of ``hacking'' or information security research in general.
\end{itemize}

Little of the value of these points in favor of vulnerability disclosure is captured directly by the firm that has been hacked. Rather, disclosure creates many positive externalities, enjoyed by the community as a whole. As such, it is likely that vulnerabilities are disclosed at a socially sub-optimal rate. 

\subsubsection*{Firm Participation in BBPs}
When starting a BBP, firms tend to set specific rules of engagement \citep{laszka2018rules}, highlighting which parts of their codebase are in scope, and committing to actions in response, for instance the level of payment and response time. When responding to individual vulnerability reports, firms need to decide whether the vulnerability is relevant and valid, and if so - how to reward the researcher. The danger in undercutting a researcher is that word travels and that other researchers may not be interested in investigating one's system. Firms have to weigh this risk against the size of their budget for the BBP.

After a vulnerability is resolved, the firm should decide whether or not to disclose the vulnerability and the interaction with the researcher. Besides appeasing the researcher, firms may signal to other researchers that they are approachable and easy to deal with. On the other hand, firms may not want to admit having these mistakes in their systems - as it may cause them to suffer reputational damage and lose esteem in the eyes of users and customers. 

\subsubsection*{Online Platforms as Market Makers for BBPs}
These tensions on both the firm and information security researcher side impose significant costs on their transactions, and likely limit the potential of decentralized BBPs in which each firm organizes its own program. Indeed, the vulnerabilities as assets are highly specific, relatively infrequent, and involve significant uncertainties. While vulnerabilities are can be categorized in general, the particular details of a vulnerability within a system and its consequences tend to be unique. A researcher may report multiple vulnerabilities to a firm, but these will likely come at irregular intervals. And finally, the process of reporting a vulnerability carries significant uncertainty for both researcher and firm. A firm may reject a vulnerability as irrelevant or undervalue it relative to the researcher's expectations, or may simply ignore the researcher, or in extreme cases consider the vulnerability report a trespass a report the researcher to the authorities \citep{goerzen2022wearing}. Firm-organized bug bounty programs that solicit reports often try to address this issue by setting rules of engagement and a menu of rewards. However, because vulnerabilities are so specific, their valuation always requires some coordination.

Transaction Cost Economics suggests that transactions with these characteristics are likely to be carried out within firms \citep{williamson1979transaction,alt2017electronic}. This likely explains the relatively slow proliferation of BBPs in the early 2000s, before the emergence of centralized platforms like HackerOne and Bugcrowd that manage bounties for many firms. These online platforms reduce information asymmetries and other frictional costs associated with the transaction of specific, infrequent, and uncertain assets. They achieve this by introducing reputation mechanisms and providing standardized procedures.

Nagle, Seamans, and Tadelis \citeyearpar{nagle2020transaction} highlight reputation mechanisms as a key mechanism by which online platforms reduce transaction costs and make markets. On HackerOne, researchers and firms have profiles that report detailed information about their activity on the platform. Statistics on a firm's responsiveness to reports, and the sizes of their bounties are public information. Firm profiles also set the scope of their assets they would pay bounties on, and often state how much they pay for vulnerabilities by severity. Though in the end the judgement of the severity of a vulnerability remains subjective, providing examples of vulnerabilities meeting different grades of severity (i.e. if a user's private information can be captured using a vulnerability) decreases the degree of specificity in the market. 

Besides listing the number of vulnerabilities they have reported to various companies, researchers collect reputation points, credits, and badges on their profile pages. These elements have value as credentials, and allow researchers to build a track record demonstrating expertise. Researcher profiles can also link to an individual's personal website and to other online platforms like LinkedIn, Twitter, or GitHub. This makes the user profile a hub through which headhunters and hiring managers can learn more about them. Such functions play an increasingly important role in digital labor markets \citep{marlow2013activity,papoutsoglou2021mining}, and allow researchers to capture ever more value from their activity on the site.

HackerOne also facilitates the market for vulnerabilities by standardizing the process by which vulnerabilities are reported and communication is carried out. Several firms participating on HackerOne actually hire HackerOne employees to manage their BBPs. HackerOne also provides general guidelines, suggests best practices, and moderates disputes. The platform also serves as a record of activity for both firms and researchers, and gives both kinds of actors the chance to learn from the activities of others.

\subsubsection*{Hypotheses}
Though the case that HackerOne facilitates the market for vulnerabilities by reducing information asymmetries and allowing researchers to capture greater value from their activity is in line with the predictions of TCE and the EMH, we now propose a series of empirical hypotheses about activity on the platform to verify these connections. Specifically, we ask if users and firms respond to information incentives, and if behaviors such as specialization and embeddedness of market participants emerge over time.

First we probe whether researchers actually move to capture labor market value of their activity on the site by connecting their profiles on HackerOne to their other avatars on the web.
\begin{quote}
\textbf{H1: More active researchers will link their profiles to their other presences on the web more frequently, especially if they have disclosed vulnerability reports.}
\end{quote}

Second, we propose that firms will provide additional signals and information about their behavior in their first interactions on the platform by disclosing more reports.
\begin{quote}
\textbf{H2: Firms will be more likely to disclose vulnerabilities early on in their history on the platform.}
\end{quote}
Finally, we suggest that researchers will tend to specialize and work more frequently with specific companies as time goes on, reflecting improved market matching owing to the proliferation of information on the platform.
\begin{quote}
\textbf{H3: Researchers will tend to focus their efforts on fewer firms over time.}
\end{quote}

\section*{Empirical Analysis}
In this section we carry out an empirical analysis of data from HackerOne. We first describe our data and its source. We present descriptive statistics about researchers, firms, and reported vulnerabilities. We then analyze the observed behavior of firms and researchers. We first observe a relationship between researcher activity and the extent to which they link to external platforms. We then investigate whether firms tend to disclose vulnerabilities earlier in their participation on the platform using a regression modeling framework. Finally we study the specialization behavior of researchers, measuring their tendency to either diversify between companies or focus on specific companies. 

\subsection*{Data}

We collect data from HackerOne, one of the largest online platforms managing BBPs. Vulnerabilities, referred to as \textit{hacks}, are the main unit of interaction on HackerOne. Tracing the stream of publicly listed hacks (around 20\% of the total vulnerabilities managed by HackerOne \citep{hackerone2019}) back in time to the launch of the site, we collect data on 124,172 vulnerabilities reported by researchers to firms from January 2014 to November 2021. We record the latest time of the interaction, the researcher and firm involved, whether the vulnerability was disclosed, and, if announced, how much money the researcher received for the report. Only 10,966 ($\approx 9\%$) of vulnerabilities were disclosed. 24,091 ($\approx 19\%$) of reported vulnerabilities lead to a visible cash payment, summing up to just over \$25 million. These cash payments were highly heterogeneous, with a mean of 1,047, a standard deviation of 2,600, and a median of 300.

Our dataset includes 14,371 researchers. We also visited each of their profile pages and recorded information about their registration date on HackerOne and whether or not they link to other platforms. We also record the free-text location provided by 3,700 researchers. Geolocating these locations using the Bing Maps API, we could infer a country-level location for 3,354 of them. The top six countries were India (704, 21\%), US (520, 16\%), UK (148, 4\%), Russia (131, 4\%), Germany (116, 3\%) and Brazil (82, 2\%). This ranking is highly similar to the one reported in the 2019 HackerOne survey \citep{hackerone2019}, in which India, the US and Russia were reported as the top three countries. We note that the geographic distribution of HackerOne researchers seems to more closely resemble the geographic distribution of remote/gig workers \citep{braesemann2021polarisation} than that of software developers \citep{wachs2022geography}.

We found 502 firms in the history of public vulnerabilities. These contain a wide range of entities, from Fortune 500 mainstays like General Motors and Starbucks to technology companies like Adobe, Facebook, and Spotify, to smaller startups. Though we refer to firms, non-business entities are also present including public sectors actors like the Department of Defense and non-profit organizations and open source software projects like Ruby.

\subsection*{Researcher Profiles}
It has been suggested that a track record of successful vulnerability reports has significant value on the information security labor market \citep{mumtaz2020security,ellis2022bounty}. To realize this value, individuals must be able to credibly link their activity on platforms like HackerOne to their other identities on the web, for instance on coding platforms like GitHub, professional platforms like LinkedIn, or social platforms like Twitter. We collected data from researcher profile pages on HackerOne, recording which other platforms they link to. In Table \ref{tab:profile_links}, we report the share of researchers linking to an external platform by the number of reports they submitted. Specifically we look for any links to personal websites, GitHub, Bugcrowd, Gitlab, Linkedin, or Twitter. We observe that while only one in three researchers in the whole population of researchers link to an external platform, nearly two in three that have made at least 10 vulnerability reports (59\%) do so. Researchers with at least 2 \textit{disclosed} reports share an external link at the same rate (59\%). 80\% of researchers with at least 10 disclosed reports have an external link.

\begin{table}
\begin{tabular}{|l|c|}
\toprule
 $\ge K$ Vuln. & Share of Researchers Linking to External Platform\\
\midrule
1 & 33\% \\
2 & 42\%\\
5 & 52\% \\
10 & 59\% \\
\bottomrule
 $\ge K$ Disclosed Vuln. & Share of Researchers Linking to External Platform\\
\midrule
1 & 33\% \\
2 & 59\%\\
5 & 71\% \\
10 & 79\% \\
\bottomrule
\end{tabular}
\caption{The share of researchers with at least $K$ credited vulnerability reports sharing links to external platforms, repeated for $K$ \textit{disclosed} (i.e. publicly visible) reports. More active researchers are more likely to link to their avatars on other platforms, and do so at a greater rate if they have more disclosed reports to their name. We count links to personal websites, GitHub, Bugcrowd, Gitlab, Linkedin, or Twitter.}
\label{tab:profile_links}
\end{table}

We observe a relationship that is perhaps not surprising: the more active a researcher is on HackerOne, the more likely they are to link to other platforms. That this likelihood sharply increases when a researcher has disclosed reports suggests that there are information effects at play. Specifically, disclosed vulnerability reports may have significant value as credentials, demonstrating in greater detail what a researcher is capable of, or for bragging rights \citep{ellis2022bounty}. Researchers seem to move to capture this extra value by linking to their other online presences. The most prolific researchers tend to make themselves more visible.

\subsection*{Disclosure}

As discussed above, disclosed vulnerability reports do not only explain the vulnerability and researcher compensation. They offer a transparent look at the history of communication between researcher and firm for a given vulnerability. Disclosed vulnerabilities show how quickly firms respond, how demanding they are for details, and how they treat the reporting researcher. Yet disclosure of vulnerabilities may have indirect costs for firms. They may suffer reputational damage, or inadvertently highlight other vulnerabilities in their systems.

Researchers may be suspicious of a firm new to BBPs or HackerOne. We hypothesized that a firm will be more likely to disclose vulnerabilities early on in their history on HackerOne to signal their quality as partners. We test this hypothesis using the following model for the binary outcome of vulnerability disclosure:

\begin{equation}\label{eq:disclosure}
    \text{Disclosed} = \beta \text{(Early Firm Vulnerability)} + \alpha_f + \gamma_y + \mu_r  + \epsilon
\end{equation}

The key variable, \textit{Early Firm Vulnerability}, is a binary indicator that takes the value 1 when a vulnerability was reported early in the firm's history on HackerOne (quantified as the first $K-th$ percentile of the firm's vulnerabilities up to the end of 2021. We apply varying thresholds to define an early vulnerability for a firm (i.e. if it is among the first 10 or 20 percent of vulnerabilities reported to a firm). A positive and statistically significant estimate of coefficient $\beta$ would indicate that firms are more likely to disclose early vulnerabilities. 

Given that firms may have heterogeneous incentives to disclose or withhold disclosure, for instance depending on their industry, we include firm fixed-effects $\alpha_f$ to control for time-invariant aspects of firms. We also control for year of the vulnerability $\gamma_y$ to control for changes in disclosure trends in general. We also include fixed-effects for the reporting researcher $\mu_r$. We cluster standard errors at the researcher-firm level. This specification controls for unchanging attributes of researchers and firms, as well as global trends - making it quite conservative.

In our primary specification we fit linear probability model using OLS rather than a logistic regression because a fair share of the firms and researchers have either 100\% or 0\% disclosure rates. A logit model discards such observations \citep{caudill1988advantage}. Our findings replicate in a logit model, which however can only be fit with firm and year fixed-effects. In our primary specifications, we also exclude data from 2021 as there is often a lag between a vulnerability's first announcement in the public HackerOne feed and its disclosure. Our results are substantively unchanged if 2021 is included.

\begin{table}
\begin{tabular}{lcccc}
\tabularnewline\midrule\midrule
Dependent Variable: & \multicolumn{4}{c}{Vulnerability Disclosed}\\
& (1) & (2) & (3) & (4)\\
 &  LPM & LPM & Logit & Logit\\
\midrule \emph{Vuln. among Firm's} &   &   &   &  \\
\\
First 10th Pctile. & 0.022$^{**}$ &    & 0.458$^{**}$ &   \\
  & (0.008) &    & (0.168) &   \\
First 20th Pctile. &    & 0.021$^{**}$ &    & 0.421$^{***}$\\
  &    & (0.007) &    & (0.134)\\
\midrule \emph{Fixed-effects} &   &   &   &  \\
Firm & Yes & Yes & Yes & Yes\\
Year & Yes & Yes & Yes & Yes\\
Researcher & Yes & Yes & -  &  -\\
\midrule \emph{Fit statistics} &   &   &   &  \\
R$^2$ & 0.52 & 0.52& - &- \\
Pseudo R$^2$ & -& -& 0.31& 0.31\\
Observations & 96,048 & 96,048 & 79,285 & 79,285\\
\midrule\midrule\multicolumn{5}{l}{\emph{Researcher-Firm-level clustered standard-errors in parentheses}}\\
\multicolumn{5}{l}{\emph{Signif. Codes: ***: 0.01, **: 0.05, *: 0.1}}\\
\end{tabular}
\caption{Models predicting whether a vulnerability reported by a researcher to a firm was publicly disclosed. In both the linear probability model (OLS) and logistic regression specifications, we observed that vulnerabilities reported early in a firm's time on the platform are more likely to be disclosed.}
\end{table}

In all four variations of our model we observe a strong effect: firms are significantly more likely to disclose vulnerability reports early in their tenure on HackerOne than later. In our preferred specification, the linear probability model with firm-year-researcher fixed effects, we estimate that early vulnerabilities are 2.2\% more likely to be disclosed. Given that only 9\% of vulnerabilities are disclosed overall, this is a large deviation. We interpret this result as evidence that firms are eager to signal how it is like to work with them earlier in their presence on the platform, when the value disseminating such information is highest. As a firm builds a long track record, they can be more selective in what vulnerabilities they choose to disclose.

\subsection*{Researcher Specialization}

Individuals and firms operating in large markets often specialize in certain sub-markets \citep{capon1988corporate}. Besides the unilateral returns researchers may gain from specialization \citep{nedelkoska2019skill,neffke2019value}, for instance in a certain class of vulnerabilities or technological stacks used by multiple firms, they likely gain from bilateral specialization \citep{di2017value}. That is to say by repeatedly interacting with the same firms. Such hybrid \citep{williamson1991comparative} or networked \citep{powell1990neither,alt2011twenty} structures often emerge in markets in which trust is important and assets remain differentiated. As vulnerability reports are highly specific and the reactions of firms difficult to predict, we expect that this is the case on HackerOne.

Such markets are characterized by social connections forged by repeated interactions \citep{uzzi1996sources}. Indeed, this notion of embeddedness of ties is known to play a crucial role in online markets with information asymmetries. In the most sensitive markets, such as those for illegal drugs or other illicit goods, embeddedness builds a sort of trust without which such markets could not function \citep{paquet2018assessing,duxbury2021network,dupont2021countering}. Embeddedness has been observed also in very traditional offline markets, such as fish markets \citep{hernandez2018trust}, in which goods are specific or have idiosyncratic quality.

We have hypothesized that such embeddedness, manifesting as repeated interactions between researchers and firms, will increase over the history of HackerOne. We suggest there are two reasons for this. The first is that highly active researchers will explore more at the beginning of their activity, focusing on the most profitable and enjoyable relationships once they have tested a variety of firms. The second is that information effects will improve initial market matching over time. Specifically, new researchers observe previous activity and disclosed reports, and so can more quickly find the right firms to work with given their skillsets and interests.

\begin{figure}
    \centering
    \includegraphics[width=0.7\textwidth]{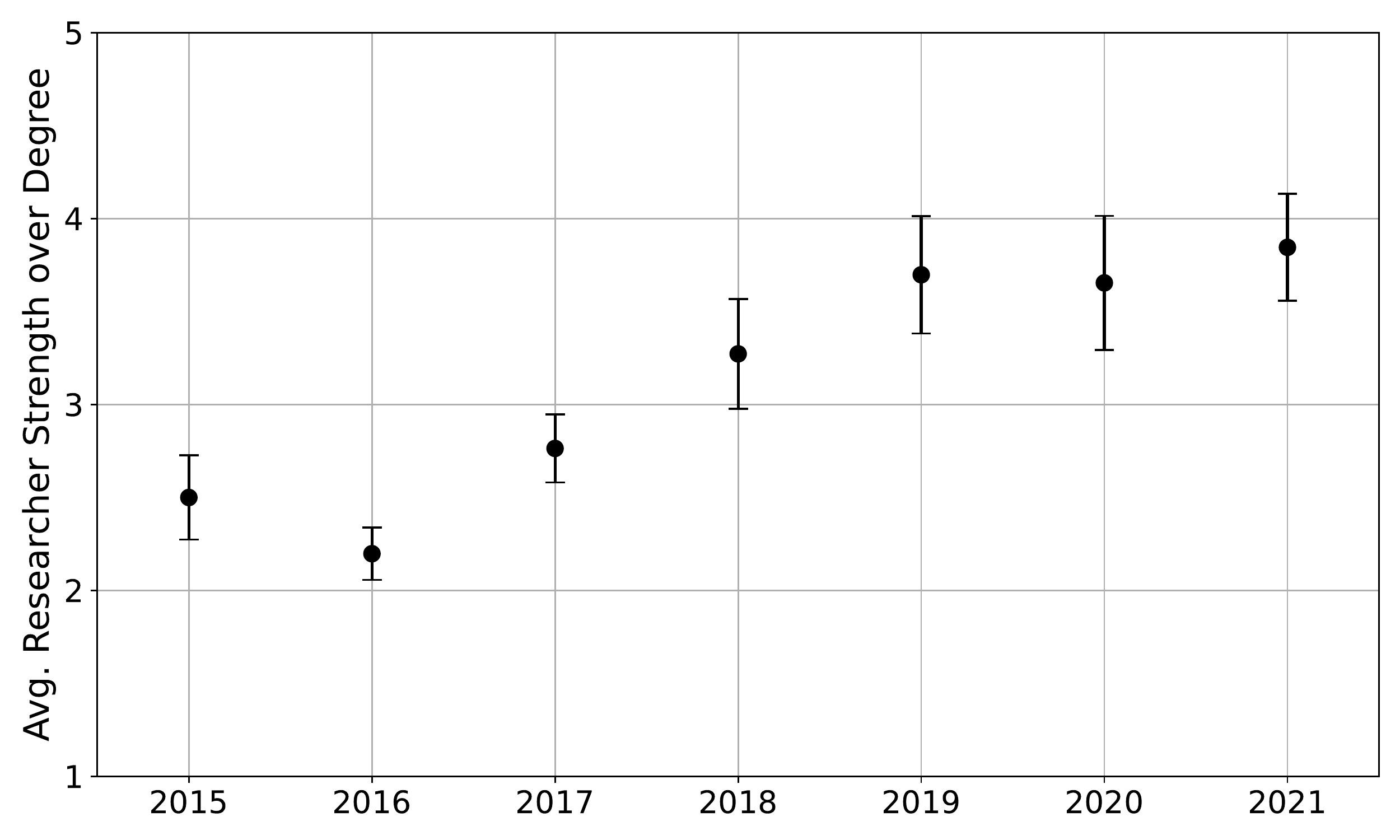}
    \caption{The average ratio of researcher strength to degree, by year. Strength is the number of vulnerabilities reported by a researcher, while degree is the unique number of firms they report vulnerabilities to. A higher strength to degree ratio indicates researchers are reporting more vulnerabilities to the same firms. We observe an increase in the relative frequency of repeated interactions over time. Error bars indicate 95\% bootstrapped confidence intervals.}
    \label{fig:strength_over_degree}
\end{figure}

We test this idea by calculating the average ratio of researcher strength to degree per year from 2015 to 2021. Borrowing terms from the networks literature, researcher strength is defined as the number of vulnerabilities they are involved in. Researcher degree is defined as the count of the unique firms they report vulnerabilities to. The ratio of strength to degree then measures the relative concentration of a researcher on a specific firm. We report the average ratio across all researchers with at least two vulnerability reports in each year, along with bootstrapped confidence intervals in Figure \ref{fig:strength_over_degree}.

We observe that the average strength to degree ratio hovers between 2 and 3 early on in the history of the site, then increases to nearly 4 by 2020/2021. Conservatively, this suggests a roughly 50\% increase in the rate at which researchers specialize on specific firms, over a relatively short time period.

To conclude our analysis we present a practical application of our observation that researchers often specialize their activity on firms. Specifically, we visualize the landscape of firms as a network in which nodes are firms and edges link firms that are often hacked by the same researcher. We note that this representation of firms in a network only makes sense if researchers tend to specialize on certain kinds of firms and repeatedly interact with the same ones. This visualization provides an overview of the how different firms are grouped together into sub-markets.

\begin{figure}
    \centering
    \includegraphics[width=\textwidth]{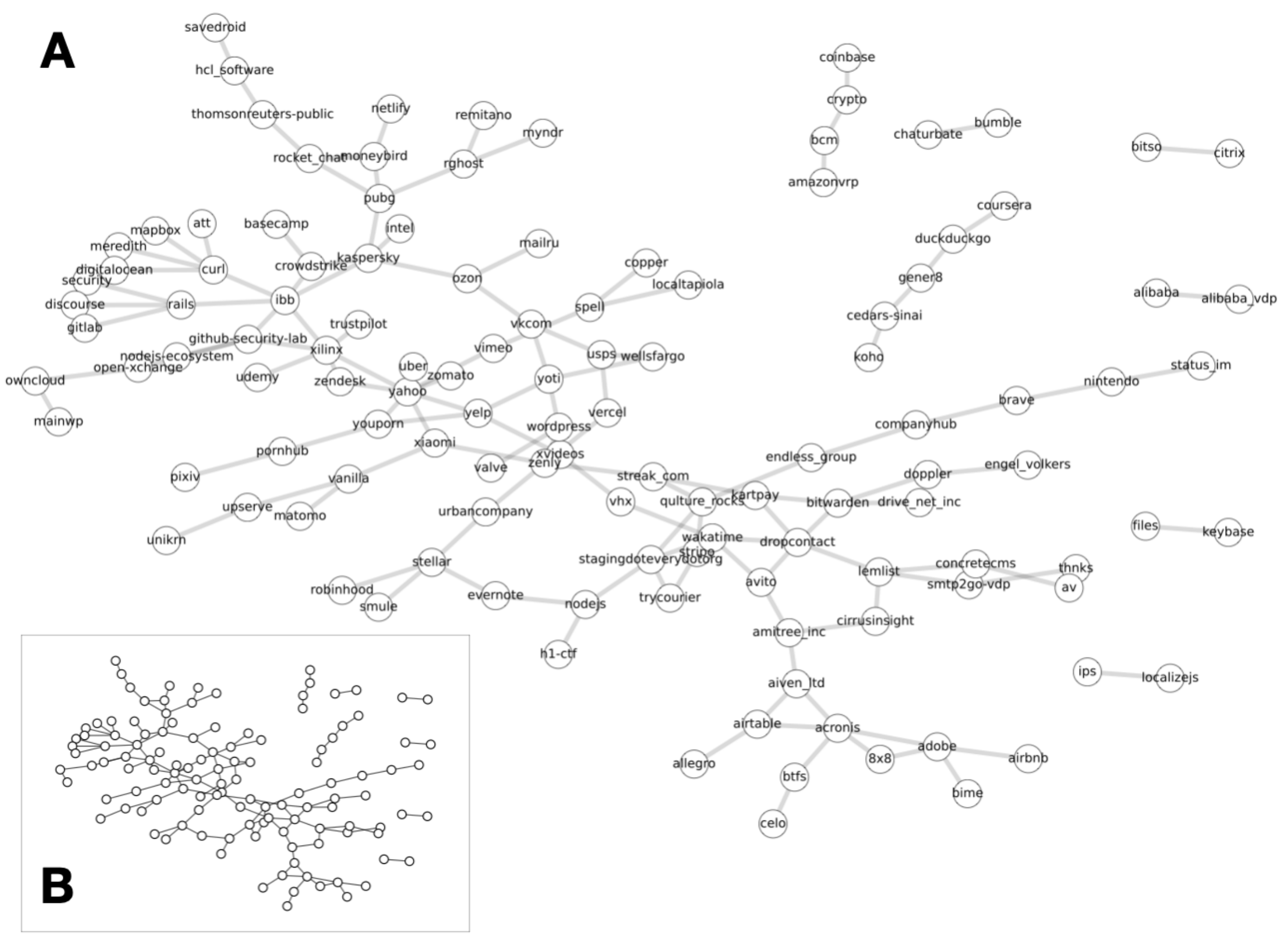}
    \caption{The network of firms on HackerOne using data from 2020. Two firms are connected if the same researchers report vulnerabilities to both. Edges are filtered for statistically significant co-occurrences of researchers (using the Noise-Correct Backbone method of Coscia and Neffke \citeyearpar{coscia2017network}. The network is plotted with labels in A), while the inset B) is the same network without labels. To some extent we observe clustering of entities in related sectors - commercial internet platforms, pornographic websites, open source software communities, and  cryptocurrency related firms.}
    \label{fig:net}
\end{figure}

To construct this network of firms we consider all vulnerabilities reported in 2020. For the sake of visualization, we remove firms with less than 10 reported vulnerabilities, and do not consider researchers submitting only one report. We connect two firms with an edge weighted by the number of researchers they have in common. We then filter the network for statistically significant co-occurrences using the Noise-Corrected Backbone method of Coscia and Neffke \citeyearpar{coscia2017network} (with the NC parameter set to 35). This method compares the observed distribution of edge weights against a hypergeometric statistical null model in which researchers and firms have the same number of interactions, but are otherwise randomized. The remaining edges represent pairs of firms which have highly significant overlap in their researchers. We visualize the network in Figure~\ref{fig:net}, dropping firms that have become isolated after the edge filtering process.

This qualitative visualization of the data serves several purposes. First, it indicates the variety of firms involved in BBPs on HackerOne, and that firms or entities in similar sectors are close in this network. For instance, pornographic websites are close to one another; as are open source software supporting entities and firms in the cryptocurrency space. It also suggests that there are significant latent similarities between certain firms that researchers are exploiting, likely relating to the technologies and systems they use. Future research could investigate how researchers navigate this landscape of firms.

\section*{Discussion}
In this paper we examined the organization of bug bounty programs, highlighting first the significant transaction costs arising from information asymmetry problems, and then studying how centralized online platforms like HackerOne reduce these problems and their costs. Doing so, these platforms make a better market for information security. They do so by enabling their participants to better signal information about themselves and to capture the value of their activity. Our work suggests that central platforms play a crucial role in the market for BBPs \citep{zhao2017devising}. Our work fits into a broader research program that demonstrates the continued relevance of the transaction cost perspective in the context of new digital economy sectors like the sharing economy and market platforms \citep{henten2016transaction,nagle2020transaction}. Our findings highlight the unique incentives and costs that researchers and firms face in this market.

We observed that researchers are more likely to link their HackerOne profiles to their other identities on the web as they are more active on the site. This observed correlation is even stronger among researchers with disclosed vulnerability reports in their records. These reports provide significant detail on the vulnerability, and act as convincing proof of a researcher's ability, akin to an example in a designer's portfolio.

On the firm side, we found that firms are significantly more likely to disclose vulnerabilities early on in their tenure on HackerOne. We interpreted this as evidence that firms wish to signal their collaborative quality to the wider crowd. Once that track record has been established, firms seem to become more conservative with what they release.

Finally we observed that over the history of HackerOne, researchers tended to become more specialized, interacting more frequently with the same firms. We interpreted this as a sign of the maturity of the market in which embedded relationships are increasingly important. We also considered an explanation involving information effects: as the history of firm activity on HackerOne grows, new researchers can more easily match with compatible firms because they can observe past firm behavior.

Despite the proliferation of BBPs, no doubt in part because of the success of platforms like Hackerone, there are still significant problems with the way BBPs work in practice \citep{ellis2022bounty,goerzen2022wearing}. For example, payment for work on HackerOne can be highly variable. A researcher can work on a vulnerability report, only to find that another researcher had already pointed out that bug. Usually these researchers get no financial rewards. For many researchers it may be difficult to generate a reliable source of income from such work. While the potential for learning and the labor market value of a strong HackerOne track record compensate researchers to some extent, such issues likely limit participation. In this way BBPs suffer many of the same problems that other gig economy jobs do \citep{graham2017digital}. Future work may consider how to optimize the design and incentives of BBPs with a view to expanding participation or smoothing out inequalities \citep{bokanyi2020understanding} on the researcher side.

The digital traces left by researchers in BBPs likely contain valuable information for the information security community. To date, we are not aware of efforts to mine data insights from disclosed vulnerabilities on platforms such as HackerOne. Such data could present a crowd intelligence complement to other sources of information such as vulnerability reports published by research institutes and governments. These mainstream sources are already collected and combined in a structured format, for instance in the SEPSES Knowledge Graph \citep{kiesling2019sepses}. Such databases form the potential basis of machine learning-based countermeasures or warning systems for vulnerabilities \citep{eckhart2020automated}. More generally, the question of how BBPs work with or complement traditional methods of security auditing remains largely unanswered.

Another promising avenue of questions about BBPs relates to entry into this market on the firm side. Information security is a high risk and sensitive area for most firms and institutions. Internal actors may rather ignore potential problems than acknowledge them, a fact reflected in widespread  underinvestment in information security among private sector firms \citep{gordon2018empirical}. Some firms may not feel comfortable engaging with semi-anonymous researchers. Many potential market participants may feel blocked from participation because they are uncomfortable with variability in the expenditures on security. This latter point may be one reason why public sector institutions are underrepresented in this space, as they are in other new economy sectors such as the sharing economy \citep{ganapati2018prospects}. This last point raises two practical questions. First, why are institutions hesitant to try BBPs, and to what extent is it because of internal organization or culture? Second, can BBPs be designed, perhaps via platforms like HackerOne, in a way that overcomes these obstacles? These questions are essential ones if we want to expand participation in BBPs, and to enjoy the resulting potential improvements in digital security.

\section*{Acknowledgements}
I would like to acknowledge and thank Sebastian Bockholt, Lukas Ochse, Edgar Weippl, Anna Weichselbraun, Sandor Juhasz, William Schueller, Gergo Toth, and Aleksandra Urman for helpful conversations that improved this manuscript.

\singlespacing
\bibliographystyle{spbasic}      
\bibliography{bibliography.bib}   

\end{document}